\documentclass[aps,prl,twocolumn,showpacs]{revtex4}  
\usepackage{bm}
\usepackage{graphicx}
\usepackage{amsmath}
\usepackage{amssymb}
\usepackage[dvips]{color}

\def\oaf{\omega^{\rm af}}
\def\otr{\omega^{\rm tr}}

\begin{document}
\title{Dynamic weakening by acoustic fluidization during stick-slip motion} 
\author{F. Giacco$^{1,2}$, L. Saggese$^{3}$, L. de Arcangelis,$^{3,4}$,  E. 
Lippiello,$^{2,4,*}$, M. Pica Ciamarra$^{5,1}$}
\affiliation{%
$^{1}$CNR--SPIN, Dept. of Physics, University of Naples ``Federico II'', 
Naples, 
Italy\\
$^{2}$Dept. of Mathematics and Physics, Second University of Naples and CNISM, 
Caserta, Italy \\
$^{3}$Dept. of Industrial and Information Engineering, Second University of 
Naples and CNISM, Aversa (CE), Italy\\
$^{4}$Kavli Institute for Theoretical Physics, University of 
California, Santa Barbara, CA 93106-4030 USA\\
$^{5}$Division of Physics and Applied Physics, School of
Physical and Mathematical Sciences, Nanyang Technological University, Singapore 
}

\email[Corresponding author: ]{eugenio.lippiello@unina2.it}

\begin{abstract}
The unexpected weakness of some faults has been attributed to the emergence of acoustic waves that promote failure by reducing the confining pressure through a mechanism known as acoustic fluidization, also proposed to explain earthquake remote triggering. Here we validate this mechanism via the numerical investigation of a granular fault model system.  We find that the stick-slip dynamics is affected only by perturbations applied at  a characteristic frequency corresponding  to  oscillations  normal to the fault,  leading to  gradual dynamical weakening  as  failure is approaching. Acoustic waves at the same frequency spontaneously emerge at the onset of failure in  absence of perturbations, supporting the relevance of acoustic fluidization in earthquake triggering.
\end{abstract}

\maketitle 

Most fault systems exhibit a resistance to shear stress much smaller than the one predicted by experiments
measuring the friction coefficient of sliding rocks~\cite{hickman91}.
A possible explanation of this unexpected observation 
is represented by
``acoustic fluidization'' (AF) that  is active 
due to the presence of crushed and ground-up rocks produced by past wearing inside the fault, 
usually defined as fault gouge~\cite{melosh79,melosh96}.
According to the AF mechanism, seismic fracture produces
elastic waves that diffuse and scatter inside the fault,  
generate a normal stress contrasting the confining one, and thus promote seismic failure. 
The same mechanism could be also activated by transient seismic waves 
generated by other earthquakes.
AF, indeed, has been also proposed to explain why
seismic activity is observed to increase, within minutes after big 
earthquakes, in areas at a distance of thousand kilometers from the mainshock 
epicenter~\cite{brodsky2006,brodsky2014,BKK00}.
Because of the large distance and the rapid response, the 
passage of seismic waves represents the most reasonable explanation for this 
remote triggering~\cite{hill93,gomberg2004,gomberg2005}. Indeed, these seismic 
waves could scatter inside a fault, effectively reducing the confining pressure and promoting failure.
To investigate this scenario, experiments have considered 
granular based models of seismic faults~\cite{johnsonjia2005,johnsonsava2008,johnson2012,vanderelst2012},
and demonstrated that acoustic perturbations
can produce lasting changes in the granular 
rheology, together with stick-slip events~\cite{johnsonsava2008,johnson2012} 
and auto-acoustic compaction~\cite{vanderelst2012}.
For instance, acoustic waves have been recently shown
to induce up to a ten--fold decrease in the 
frictional strength~\cite{xia2011pre,xiamarone2013} of granular materials.
Consistently, numerical investigations have also demonstrated that vibrations can  advance the time of slip instability~\cite{griffa2013,3ddegriffa2014}.
Other studies have identified a frequency regime leading to
friction reduction~\cite{capozzaprl2009,giaccopre} caused  the detachment of the particles from the vibrating confining planes and not related to the fluidization of the granular bed.

Here we validate the AF scenario through the numerical investigation of a
model system. First we    show that perturbations at a characteristic resonant frequency are able to activate acoustic modes inside the fault, 
promoting the entire system fluidization. Then we clarify that these modes are relevant in the AF scenario as they  spontaneously emerge at the onset of slip instabilities, in the absence of perturbations.

\begin{figure}[t!]
\includegraphics*[scale=0.33]{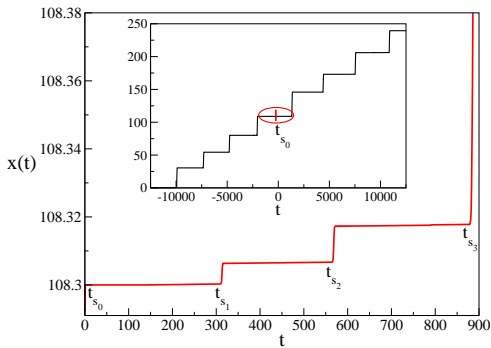}
\caption{(color online)
Time evolution of the top plate position. The inset shows the evolution 
in a large time interval,
while the main panel focuses on a shorter time interval following time $t_{s_0}=0$.
In this shorter interval we observe three small slips, at time $t_{s_0}$, $t_{s_1}=321$ and $t_{s_2}=576$,
and a large one at time $t_{s_3}=890$.
\label{fig:sslip}
}
\end{figure}

We study the AF mechanism through the numerical investigation of a granular
 fault model that reproduces
the main statistical features of real earthquake 
occurrence~\cite{prl2010,epl2011}.
The model consists of $N = 1000$ spherical grains of mass $m$
and diameter $d$, representing the fault gouge, 
confined between two rough rigid layers of size $L_x \times 
L_y = 20\,d \times 5\,d$, at  constant pressure $P_0$. 
Grains interact through a normal viscoelastic interaction
and a tangential frictional one~\cite{prl2010,epl2011}.
Periodic boundary conditions are imposed along $x$ and $y$.
A stick slip dynamics is induced by driving the system via a spring mechanism.
Specifically, one extreme of the spring is attached to the top plate, while the other moves along $x$ at constant velocity $v_d$.
Accordingly, if the plate does not move the shear stress $\sigma=\sigma_{xy}$ increases at a rate $k_d v_d/L_x L_y$,
where $k_d$ is the spring elastic constant.
We measure the mass in units of $m$, the lengths in units of $d$ and time in 
units of $\sqrt{m/k_{d}}$. The confining pressure is $P_{0}= {k_d/d}$,  $v_d=0.01\, d/\sqrt{m/k_{d}}$ and the temporal integration 
step of the equations of motion is $5\cdot 10^{-3}\,\sqrt{m/k_{d}}$.  
For these values of the parameters, the fault width slightly fluctuates around 
$W \simeq 10\,d$ and the system exhibits a stick-slip 
motion~\cite{modphys2009,prl2010,epl2011}, as shown in Fig.~\ref{fig:sslip}.
We identify slips imposing a threshold $10^{-4}$ on the top plate velocity. 
We have previously shown that the slip size distribution of this model
has a power law regime in agreement with the Gutenberg-Richter law observed for earthquakes 
\cite{prl2010,epl2011}, and a bump at large slips related to the system size.

\begin{figure}[t!]
\includegraphics*[scale=0.5]{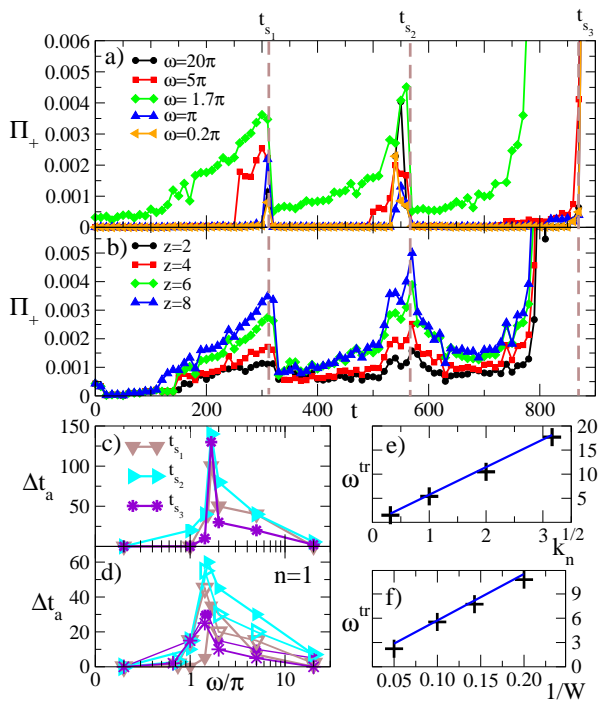}
\caption{
(color online) (a) Time dependence of the system's response $\Pi_{+}(t,W-1)$ to perturbations 
increasing the confining pressure, at different frequencies. (b) The response $\Pi_{+}(t,z)$ for grains  belonging to different vertical positions $z$. 
(c) Frequency dependence of the advance time $\Delta t_a$ 
of the slips occurring at times $t_{s_1}$, 
$t_{s_2}$ and $t_{s_3}$, induced by perturbation  $P_{+}(t)$.
(d) As in panel c, for single pulse perturbations increasing the
pressure (full symbols) or decreasing the shear stress (open symbols).
Panels (e--f) show that the characteristic triggering frequency scales
as $\otr \propto k_n^{1/2}/W$, with 
 $k_n$ grain stiffness and $W$ fault width.}
\label{fig:figu2}
\end{figure}

In the following we focus on the temporal 
window $[t_{s_0}:t_{s_3}]$ illustrated in Fig.~\ref{fig:sslip}, but analogous results are  
observed in other temporal intervals.
In this time window we observe three small slips, at times $t_{s_0}=0, 
t_{s_1}$ and $t_{s_2}$, where the displacement of the top plate $\delta x(t_{s_j})<0.1d$ 
followed by a large slip, at time $t_{s_3}$ with  $\delta x(t_{s_3}) \simeq L_x$. 
Both small and large slips involve the rearrangement of all grains inside the fault. 
In particular, the velocity profile during all slips  is compatible with a laminar flow indicating
that slip instabilities correspond to a transition from a jammed solid-like to an unjammed fluid-like configuration.    
Our goal is to understand whether 
an acoustic perturbation can cause this fluidization, and whether in  absence of 
perturbations spontaneous acoustic emissions occur at slip instabilities.
To mimic an acoustic perturbation resulting from an incoming seismic wave,
at each time $t$ we consider a replica of the unperturbed system.
Each replica is then perturbed 
by means of a series of $n$ sinusoidal stress pulses of total duration $\tau$, in absence of the external drive ($v_{d}=0$).
Specifically, either we force the shear stress
to vary by $\sigma_\pm(t_p,t)= \pm \frac{\alpha 
\sigma(t)}{2}\left[1-\sin(\frac{\pi}{2}+\omega (t_p-t)) \right ]$, 
or the confining pressure by $P_{\pm}(t_p,t)=  \pm \frac{\alpha 
P_{0}}{2}\left[1-\sin(\frac{\pi}{2}+\omega(t_p-t)) \right ]$, 
where $t$ indicates that we are considering the replica taken at time $t$, whereas $t_p$ refers to the time evolution of the perturbed system. 
We also found that purely sinusoidal perturbations lead to analogous results.

We fix the  duration of each perturbation to $\tau=10$ and consider frequencies $\omega$ leading to $n = \frac{\tau \omega}{2\pi} \in [1,10^3]$ pulses, 
restricting to the linear response regime $\alpha \ll 1$. 
In this regime the perturbing pressure is much smaller than the confining one, and it is not able to induce the detachment of the grains from the confining plates. 
The situation is different from the procedure applying a mechanical vibration to the the bottom substrate as in the study of ref.~\cite{capozzaprl2009,giaccopre}. 
In this case, indeed, for a given amplitude of the applied vibration there is a range of frequency leading to the detachment from the bottom substrate. 
We also note that the perturbations $\sigma_+$ and $P_{-}$ correspond 
to an increase of the shear stress and to a reduction of the confining pressure, respectively,
and are thus expected to facilitate failure.  Conversely the perturbations
$\sigma_-$ and $P_{+}$ should inhibit failure.

The perturbation leads to a global rearrangement of grains inside the system.
We thus quantify the effect of the perturbation by means of the displacement 
$\Delta x(t,z)=x_\alpha(t+\tau,z)-x_0(t+\tau,z)$,
where $x_{\alpha}(t+\tau,z)$ is the average position  
of all grains with vertical position $\in [zd,zd+d)$ after
the perturbation has been applied, and $x_0(t+\tau,z) \simeq x_0(t,z)$ is their average  unperturbed position.
We use these displacements to estimate the frictional weakening in response to the 
the perturbations $P_{\pm}$ and $\sigma_{\pm}$ via 
the parameters $\Pi_\pm(t,z)= \Delta x(t,z)/\alpha P_0$ and $\Sigma_\pm(t,z) = \Delta 
x(t)/\alpha \sigma(t,z)$. In the following, we mainly focus on the top plate 
response  $\Pi_\pm(t,W-1)$ or  $\Sigma_\pm(t,W-1)$ and 
report results obtained
in the linear response regime ($\alpha < 0.05$), where 
$P_{\pm}$ and $\sigma_{\pm}$ are $\alpha$ independent.

We first consider the response to an increase
of the confining pressure, expecting to observe
$\Pi_{+} \simeq 0$ since a pressure increase is
supposed to keep a system in a jammed state. 
Indeed, Fig.~\ref{fig:figu2}a shows that for most frequencies $\Pi_{+}\simeq 0$,
as long as $t$ is not very close to a slip occurrence time.
However, 
for $\omega=1.7\pi$ the response is
significantly different than zero at all times.
To investigate the frequency dependence of the response, 
we observe that an external perturbation applied at a time $t < t_{s}$ induces a displacement   $\Delta x(t,W-1)$  which is smaller but comparable to  the one  observed in the unperturbed system when the failure occurs at time $t_s$, $\delta x(t_s)$.   The induced displacement $\Delta x(t,W-1)$ becomes larger and larger as $t$ approaches $t_s$  and can be used to  define the `advance time', $\Delta t_a$,
 by the condition $\Delta x(t_s-\Delta t_a,W-1) = \beta \delta x(t_s)$. The advance time    depends on both $\alpha$ and $\beta$, increasing with $\alpha$  and decreasing with $\beta$.     
In Fig.~\ref {fig:figu2}b we plot the frequency dependence of  $\Delta t_a$ for  $\alpha = 0.02$ and  $\beta = 0.2$ for all the considered slips. 
The behavior of $\Delta t_a$ indicates that compressive perturbations trigger failure when their frequency falls in a given range, the triggering
being most effective at a particular frequency, $\otr=1.7\pi$. For larger values of $\beta$,  $\Delta t_a$ converges to a delta function centered in $\omega^{tr}$. 
Fig.~\ref{fig:figu2}c shows that analogous results are obtained 
when a single pulse is applied, and thus clarifies that  triggering
is related to the frequency of the perturbation, not to its duration.
Moreover the perturbation does not only cause the displacement of the top plate but involves a non-local rearrangement of all granular layers. 
This is clearly enlightened by  Fig.~\ref {fig:figu2}b where we plot the response  $\Pi_+(t,z)$, at frequency $\omega^{tr}$ for different values of $z$.  
We observe that the external perturbation 
induces the displacement of all  system layers, with a slip profile consistent with a laminar flow.
Similar behavior is observed for the other temporal windows considered, with the response  $\Pi_+$ monotonically increasing with 
the slip amplitude $\delta x(t_{s})$.   
We also note that the magnitude of the system's response is affected by the viscoelastic nature of the interaction between the grain,
and increases (decreases) if this interaction becomes less (more) dissipative.

\begin{figure}[t!]
\includegraphics*[scale=0.4]{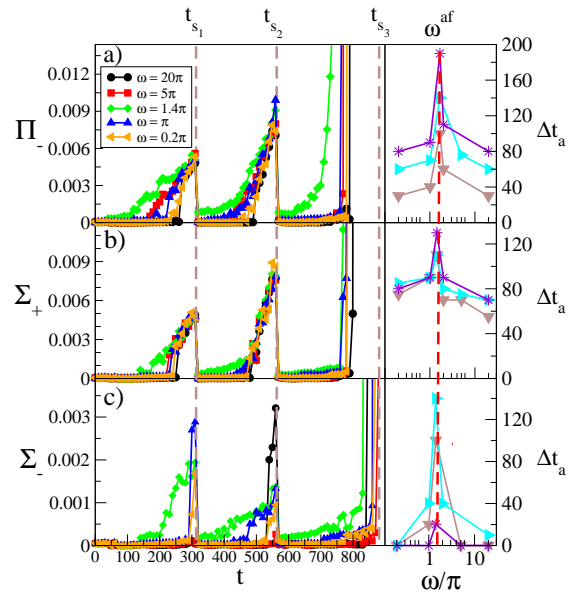}
\caption{ (color online) The left column illustrates the  response 
of the top plate
to perturbations
decreasing the confining pressure (a), increasing the shear stress (b),
or decreasing the shear stress (c). 
The right column illustrates the frequency dependence of the advance time,
which is peaked around the acoustic fluidization frequency $\oaf\simeq1.5\pi$. Symbols are as in Fig. 2b. 
}
\label{fig:figu3}
\end{figure}

We interpret the above results in terms of the fluidization 
induced by the presence of acoustic waves scattering into the system. 
These waves propagate with velocity 
$v_a=\sqrt{M/\rho}$, where $M$
is the $P$-wave modulus, and 
thus need a time $T_a=2W/v_a$ to reach the
bottom plate and return to the top. For
a single grain under a hydrostatic pressure $M\simeq k_{n}/6d$,  
and using $\rho \simeq m N/(L_x L_y W)$ the typical AF resonant frequency is
$\oaf= 2\pi/T_a = (\pi/W)\sqrt{k_n/(6 d \rho)}$. 
By performing simulations with different grain stiffnesses and system
widths, we have verified that the triggering frequency
is in agreement with $\oaf$,
as illustrated in Fig.~\ref{fig:figu2}d--e.
This proves that the AF mechanism is
at work in the response of our system to the considered perturbation.
The same conclusion is reached investigating the response to
perturbations $\Pi_{-}$ decreasing the confining
pressure, as summarized in Fig.~\ref{fig:figu3}a. 
Since reducing the pressure induces failure,
in this case the system is expected to be more sensitive to the perturbation,
and therefore we do observe $\Pi_- > \Pi_+$ 
and a not-negligible response at all times.
The frequency dependence of the advance time
clarifies that also in this case 
the system is more susceptible
to perturbations with a frequency close to $\oaf$, in
agreement with the AF scenario.

We now analyze the response to perturbations in the shear stress,
$\Sigma_{\pm}(t,W-1)$ (Fig.~\ref{fig:figu3}b--c). The perturbation $\sigma_{+}$ increases 
the shear stress, and it is therefore equivalent to a reduction of the time to 
the next slip. 
It is possible to show, indeed, that $\Sigma_{+}$ is larger for longer  
perturbation durations $\tau$. 
For a fixed $\tau$, as in Fig.~\ref{fig:figu3}b, a weak dependence on $\omega$ can be still observed with a larger response for $\otr=1.4\pi$.
Conversely, the perturbation $\sigma_-$ should inhibit the top plate 
displacement and therefore we would expect  $\Sigma_{-}\simeq 0$. This is the 
case except at $\omega^{tr}=1.4\pi$ where $\Sigma_{-}$ is larger than 
zero in a wide temporal range.
Note that the characteristic frequency observed in the response to 
perturbations in the shear stress is very close to that
characterizing the response to perturbations applied to the pressure,
and the two are found to scale in the same way with the system
width and the grain stiffness. 
Overall, these results indicate that AF can be triggered by
perturbations applied along any direction, and provide a possible 
explanation of  triggering caused by transient seismic waves  regardless 
the fault orientation.
Nevertheless, since seismic waves from remote earthquakes present small frequency signals ($\omega \lesssim 1$ Hz),  assuming $v_a \sim 1$ km/sec,
AF represents a realistic mechanism only for fault widths $W \gtrsim 10^3$ m, 
much larger than typical experimental values~\cite{melosh96,xiamarone2013}. 
The AF scenario can be still recovered if 
seismic waves are  able to excite a local source of high frequency energy or 
if seismic wave velocity abruptly decreases entering the fault granular gouge.
This velocity reduction can be attributed to spatial heterogeneity of the granular medium as measured in experiments with glass beads~\cite{andreotti}.

\begin{figure}[t!]
\includegraphics*[width=1.1\linewidth,height=0.75\linewidth]{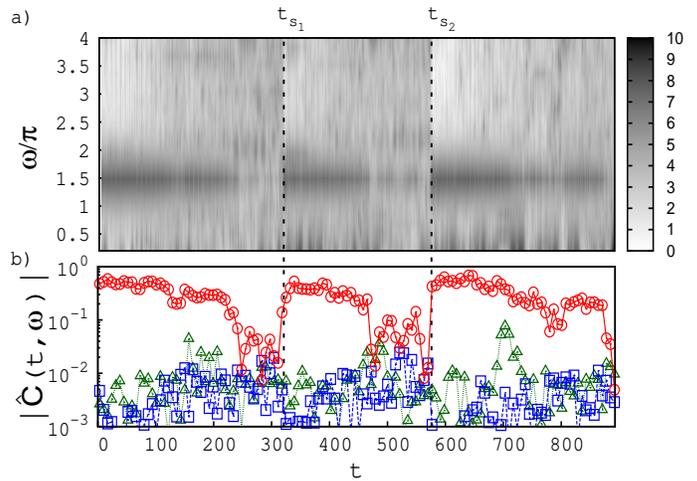}
\caption{(color online) (a) The map of the logarithm of the power spectral
density, $|\hat C(t,\omega)|$ as function of $t$ and $\omega$. (b) Time dependence of the power spectral
density at three different frequencies, $\omega\simeq \pi$ (squares), 
$\omega=\oaf\simeq 1.5\pi$ (circles) and $\omega\simeq 2\pi$ (triangles). 
The dashed vertical lines indicate the slip occurrence time $t_{s_1}$ and $t_{s_2}$.}
\label{fig:matricespettro}
\end{figure}

Having clarified the relevance of  AF 
in the response of the system to external perturbations, we now 
show that acoustic emissions can spontaneously appear and weaken a fault, 
 inducing its failure, as suggested in ref.~\cite{melosh79}. 
A similar mechanism has been used to
rationalize the compaction of sand grains under shear~\cite{vanderelst2012}.
We test this hypothesis by investigating whereas, in the unperturbed system, 
changes in the features of particle motion suggest the emergence of acoustic 
waves on approaching failure. 
To do so, at each time $t$, we create a replica of the system 
decoupled from the external drive ($v_d=0$) and follow its spontaneous 
relaxation  in the subsequent time interval.
We evaluate  the time dependence of the power spectral density $\hat 
C(t,\omega)$ obtained from  the autocorrelation function of the particle 
velocities $\vec v_{i}$, 
\[ C(t,t')= \frac{\sum_{i=1}^{N} \vec v_{i}(t)\cdot\vec 
v_{i}(t')}{\sum_{i=1}^{N} \vec v_{i}(t)\cdot \vec v_{i}(t)}. \]  
We present in Fig.~\ref{fig:matricespettro}a the map of $\log(|\hat 
C(t,\omega)|)$ for different values of  $\omega$ 
and $t\in[0,t_{s_3}-10]$ \footnote{The function $\log(|\hat C(t,\omega)|)$  cannot be evaluated in the temporal period immediately preceding the largest slip ($t \leq t_{s_3}$) since during the spontaneous relaxation, even in absence of  external drive,  the system reaches the slip instability}.
The dashed vertical black lines indicate the slip 
occurrence times  $t_{s_1}$ and $t_{s_2}$. Fig.~\ref{fig:matricespettro}a 
shows that oscillations at the characteristic frequency appear at the onset of each slip. Their amplitude then decreases roughly exponentially in time, as 
 observed in Fig.~\ref{fig:matricespettro}b. This figure also shows that oscillations at other frequencies are essentially unaffected by the slips.
This behavior is systematically observed in other slip sequences.

\begin{figure}[t!]
\includegraphics*[scale=0.47]{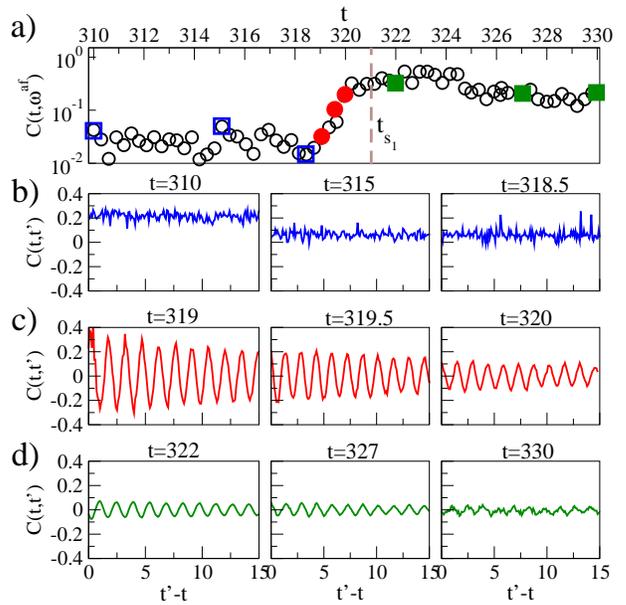}
\caption{(color online)  (Panel a) Time dependence of the power spectral density
at the characteristic frequency $|\hat C(t, \oaf)|$, in a temporal
interval centered at the slip occurrence time $t_{s_1}$. 
The bottom panels illustrate the temporal evolution of 
$C(t,t')$  evaluated at different times $t$ indicated in panel (a) as open blue squares  for $t_{s_1}-t >2$ (panel b), filled red circles $t_{s_1}-t <2$ (panel c) and filled green squares $t-t_{s_1} >2$ (panel d).}
\label{fig:fig5}
\end{figure}
Figure~\ref{fig:matricespettro} is consistent with a scenario where, as soon as acoustic oscillations 
spontaneously form inside the system, the confining pressure is reduced inducing 
slip occurrence. To investigate this hypothesis, in Fig.~\ref{fig:fig5} we focus on the 
behavior of $|\hat C(t,\oaf)|$ in a temporal period centered at the  
slip occurrence time. In the figure we also plot the evolution of $C(t,t')$ 
as function of $t'-t$ for nine different values of $t$.

In temporal intervals distant from the slip occurrence times 
$C(t,t')$ is structureless,  leading to $|\hat C(t,\oaf)| \simeq 0$. 
The same behavior is observed up to times $t_s-t >2$ before the slip. Interestingly, as soon as the slip is approaching  ($t_s-t <2$), 
oscillations at the characteristic frequency $\oaf$ appear in $C(t,t')$
and indeed $|\hat C(t,\oaf)|$ drastically increases. 
These oscillations are still present even in temporal periods after the slip, but their
amplitude decreases in time.
This result evidences that spontaneous oscillations appear at  
$\oaf$ just at the onset of the slip, favoring system failure. 
The same pattern is recovered for another slip at time $t_{s_2}$.

Summarizing, our results strongly support the validity of the AF scenario
by proving that the system is susceptible to external perturbations with a 
characteristic frequency, and that in  absence of perturbations
acoustic vibrations at this characteristic frequency spontaneously emerge 
at the onset of failure. Accordingly, these oscillations  are able to cause a 
slip instability regardless of their origin.

E.L. and L.d.A. acknowledge financial support from the National Science Foundation 
under Grant No. NSF PHY11-25915.
L.S. is supported by the MASTRI EXCELLENCE NETWORK (CUP B25B09000010007) of the Campania region.
M.P.C. acknowledges financial support from the CNR-NTU joint laboratory for
`Amorphous Materials for Energy harvesting applications'.

\bibliography{biblio}{}

\begin{thebibliography}{23}
\expandafter\ifx\csname natexlab\endcsname\relax\def\natexlab#1{#1}\fi
\expandafter\ifx\csname bibnamefont\endcsname\relax
  \def\bibnamefont#1{#1}\fi
\expandafter\ifx\csname bibfnamefont\endcsname\relax
  \def\bibfnamefont#1{#1}\fi
\expandafter\ifx\csname citenamefont\endcsname\relax
  \def\citenamefont#1{#1}\fi
\expandafter\ifx\csname url\endcsname\relax
  \def\url#1{\texttt{#1}}\fi
\expandafter\ifx\csname urlprefix\endcsname\relax\def\urlprefix{URL }\fi
\providecommand{\bibinfo}[2]{#2}
\providecommand{\eprint}[2][]{\url{#2}}

\bibitem[{\citenamefont{Hickman}(1991)}]{hickman91}
\bibinfo{author}{\bibfnamefont{S.~H.} \bibnamefont{Hickman}},
  \bibinfo{journal}{Rev. Geophys.} \textbf{\bibinfo{volume}{29}},
  \bibinfo{pages}{759} (\bibinfo{year}{1991}).

\bibitem[{\citenamefont{Melosh}(1979)}]{melosh79}
\bibinfo{author}{\bibfnamefont{H.}~\bibnamefont{Melosh}}, \bibinfo{journal}{J.
  Geophys. Res.} \textbf{\bibinfo{volume}{84}}, \bibinfo{pages}{7513}
  (\bibinfo{year}{1979}).

\bibitem[{\citenamefont{Melosh}(1996)}]{melosh96}
\bibinfo{author}{\bibfnamefont{H.}~\bibnamefont{Melosh}},
  \bibinfo{journal}{Nature} \textbf{\bibinfo{volume}{379}},
  \bibinfo{pages}{601} (\bibinfo{year}{1996}).

\bibitem[{\citenamefont{Felzer and Brodsky}(2006)}]{brodsky2006}
\bibinfo{author}{\bibfnamefont{K.~R.} \bibnamefont{Felzer}} \bibnamefont{and}
  \bibinfo{author}{\bibfnamefont{E.~E.} \bibnamefont{Brodsky}},
  \bibinfo{journal}{Nature} \textbf{\bibinfo{volume}{441}},
  \bibinfo{pages}{735} (\bibinfo{year}{2006}).

\bibitem[{\citenamefont{Brodsky and van~der Elst}(2014)}]{brodsky2014}
\bibinfo{author}{\bibfnamefont{E.~E.} \bibnamefont{Brodsky}} \bibnamefont{and}
  \bibinfo{author}{\bibfnamefont{J.~N.} \bibnamefont{van~der Elst}},
  \bibinfo{journal}{Annu. Rev. Earth Planet. Sci.}
  \textbf{\bibinfo{volume}{42}}, \bibinfo{pages}{317} (\bibinfo{year}{2014}).

\bibitem[{\citenamefont{Brodsky et~al.}(2000)\citenamefont{Brodsky, Karakostas,
  and Kanamori}}]{BKK00}
\bibinfo{author}{\bibfnamefont{E.~E.} \bibnamefont{Brodsky}},
  \bibinfo{author}{\bibfnamefont{V.}~\bibnamefont{Karakostas}},
  \bibnamefont{and} \bibinfo{author}{\bibfnamefont{H.}~\bibnamefont{Kanamori}},
  \bibinfo{journal}{Geophys. Res. Lett.} \textbf{\bibinfo{volume}{27}},
  \bibinfo{pages}{2741} (\bibinfo{year}{2000}).

\bibitem[{\citenamefont{Hill et~al.}(1993)\citenamefont{Hill, Reasenberg,
  Michael, Arabaz, Beroza, Brumbaugh, Brune, Castro, Davis, dePolo
  et~al.}}]{hill93}
\bibinfo{author}{\bibfnamefont{D.~P.} \bibnamefont{Hill}},
  \bibinfo{author}{\bibfnamefont{P.~A.} \bibnamefont{Reasenberg}},
  \bibinfo{author}{\bibfnamefont{A.}~\bibnamefont{Michael}},
  \bibinfo{author}{\bibfnamefont{W.~J.} \bibnamefont{Arabaz}},
  \bibinfo{author}{\bibfnamefont{G.}~\bibnamefont{Beroza}},
  \bibinfo{author}{\bibfnamefont{D.}~\bibnamefont{Brumbaugh}},
  \bibinfo{author}{\bibfnamefont{J.~N.} \bibnamefont{Brune}},
  \bibinfo{author}{\bibfnamefont{R.}~\bibnamefont{Castro}},
  \bibinfo{author}{\bibfnamefont{S.}~\bibnamefont{Davis}},
  \bibinfo{author}{\bibfnamefont{D.}~\bibnamefont{dePolo}},
  \bibnamefont{et~al.}, \bibinfo{journal}{Science}
  \textbf{\bibinfo{volume}{260}}, \bibinfo{pages}{1617} (\bibinfo{year}{1993}).

\bibitem[{\citenamefont{Gomberg et~al.}(2004)\citenamefont{Gomberg, Bodin,
  Larson, and Dragert}}]{gomberg2004}
\bibinfo{author}{\bibfnamefont{J.}~\bibnamefont{Gomberg}},
  \bibinfo{author}{\bibfnamefont{P.}~\bibnamefont{Bodin}},
  \bibinfo{author}{\bibfnamefont{K.}~\bibnamefont{Larson}}, \bibnamefont{and}
  \bibinfo{author}{\bibfnamefont{H.}~\bibnamefont{Dragert}},
  \bibinfo{journal}{Nature} \textbf{\bibinfo{volume}{427}},
  \bibinfo{pages}{621} (\bibinfo{year}{2004}).

\bibitem[{\citenamefont{Gomberg and Johnson}(2005)}]{gomberg2005}
\bibinfo{author}{\bibfnamefont{J.}~\bibnamefont{Gomberg}} \bibnamefont{and}
  \bibinfo{author}{\bibfnamefont{P.~A.} \bibnamefont{Johnson}},
  \bibinfo{journal}{Nature} \textbf{\bibinfo{volume}{437}},
  \bibinfo{pages}{830} (\bibinfo{year}{2005}).

\bibitem[{\citenamefont{Johnson and Jia}(2005)}]{johnsonjia2005}
\bibinfo{author}{\bibfnamefont{P.~A.} \bibnamefont{Johnson}} \bibnamefont{and}
  \bibinfo{author}{\bibfnamefont{X.}~\bibnamefont{Jia}},
  \bibinfo{journal}{Nature} \textbf{\bibinfo{volume}{437}},
  \bibinfo{pages}{871} (\bibinfo{year}{2005}).

\bibitem[{\citenamefont{Johnson et~al.}(2008)\citenamefont{Johnson, Savage,
  Knuth, Gomberg, and Marone}}]{johnsonsava2008}
\bibinfo{author}{\bibfnamefont{P.~A.} \bibnamefont{Johnson}},
  \bibinfo{author}{\bibfnamefont{H.}~\bibnamefont{Savage}},
  \bibinfo{author}{\bibfnamefont{M.}~\bibnamefont{Knuth}},
  \bibinfo{author}{\bibfnamefont{J.}~\bibnamefont{Gomberg}}, \bibnamefont{and}
  \bibinfo{author}{\bibfnamefont{C.}~\bibnamefont{Marone}},
  \bibinfo{journal}{Nature} \textbf{\bibinfo{volume}{451}}, \bibinfo{pages}{57}
  (\bibinfo{year}{2008}).

\bibitem[{\citenamefont{Johnson et~al.}(2012)\citenamefont{Johnson, Carpenter,
  Knuth, Kaproth, Bas, Daub, and Marone}}]{johnson2012}
\bibinfo{author}{\bibfnamefont{P.~A.} \bibnamefont{Johnson}},
  \bibinfo{author}{\bibfnamefont{B.~M.} \bibnamefont{Carpenter}},
  \bibinfo{author}{\bibfnamefont{M.}~\bibnamefont{Knuth}},
  \bibinfo{author}{\bibfnamefont{B.~M.} \bibnamefont{Kaproth}},
  \bibinfo{author}{\bibfnamefont{P.-Y.~L.} \bibnamefont{Bas}},
  \bibinfo{author}{\bibfnamefont{E.~G.} \bibnamefont{Daub}}, \bibnamefont{and}
  \bibinfo{author}{\bibfnamefont{C.}~\bibnamefont{Marone}},
  \bibinfo{journal}{J. Geophys. Res.} \textbf{\bibinfo{volume}{117}},
  \bibinfo{pages}{B04310} (\bibinfo{year}{2012}).

\bibitem[{\citenamefont{van~der Elst et~al.}(2012)\citenamefont{van~der Elst,
  Brodsky, Bas, and Johnson}}]{vanderelst2012}
\bibinfo{author}{\bibfnamefont{J.~N.} \bibnamefont{van~der Elst}},
  \bibinfo{author}{\bibfnamefont{E.~E.} \bibnamefont{Brodsky}},
  \bibinfo{author}{\bibfnamefont{P.~L.} \bibnamefont{Bas}}, \bibnamefont{and}
  \bibinfo{author}{\bibfnamefont{P.~A.} \bibnamefont{Johnson}},
  \bibinfo{journal}{J. Geophys. Res.} \textbf{\bibinfo{volume}{117}},
  \bibinfo{pages}{B09314} (\bibinfo{year}{2012}).

\bibitem[{\citenamefont{Jia et~al.}(2011)\citenamefont{Jia, Brunet, and
  Laurent}}]{xia2011pre}
\bibinfo{author}{\bibfnamefont{X.}~\bibnamefont{Jia}},
  \bibinfo{author}{\bibfnamefont{T.}~\bibnamefont{Brunet}}, \bibnamefont{and}
  \bibinfo{author}{\bibfnamefont{J.}~\bibnamefont{Laurent}},
  \bibinfo{journal}{Phys. Rev. E} \textbf{\bibinfo{volume}{84}},
  \bibinfo{pages}{020301 (R)} (\bibinfo{year}{2011}).

\bibitem[{\citenamefont{Xia et~al.}(2013)\citenamefont{Xia, Huang, and
  Marone}}]{xiamarone2013}
\bibinfo{author}{\bibfnamefont{K.}~\bibnamefont{Xia}},
  \bibinfo{author}{\bibfnamefont{S.}~\bibnamefont{Huang}}, \bibnamefont{and}
  \bibinfo{author}{\bibfnamefont{C.}~\bibnamefont{Marone}},
  \bibinfo{journal}{G3} \textbf{\bibinfo{volume}{14}}, \bibinfo{pages}{1012}
  (\bibinfo{year}{2013}).

\bibitem[{\citenamefont{Griffa et~al.}(2013)\citenamefont{Griffa, Ferdowsi,
  Daub, Guyer, Johnson, Marone, and Carmeliet}}]{griffa2013}
\bibinfo{author}{\bibfnamefont{M.}~\bibnamefont{Griffa}},
  \bibinfo{author}{\bibfnamefont{B.}~\bibnamefont{Ferdowsi}},
  \bibinfo{author}{\bibfnamefont{E.~G.} \bibnamefont{Daub}},
  \bibinfo{author}{\bibfnamefont{R.~A.} \bibnamefont{Guyer}},
  \bibinfo{author}{\bibfnamefont{P.~A.} \bibnamefont{Johnson}},
  \bibinfo{author}{\bibfnamefont{C.}~\bibnamefont{Marone}}, \bibnamefont{and}
  \bibinfo{author}{\bibfnamefont{J.}~\bibnamefont{Carmeliet}},
  \bibinfo{journal}{Phys. Rev. E.} \textbf{\bibinfo{volume}{87}},
  \bibinfo{pages}{012205} (\bibinfo{year}{2013}).

\bibitem[{\citenamefont{Ferdowsi et~al.}(2014)\citenamefont{Ferdowsi, Griffa,
  Guyer, Johnson, Marone, and Carmeliet}}]{3ddegriffa2014}
\bibinfo{author}{\bibfnamefont{B.}~\bibnamefont{Ferdowsi}},
  \bibinfo{author}{\bibfnamefont{M.}~\bibnamefont{Griffa}},
  \bibinfo{author}{\bibfnamefont{R.~A.} \bibnamefont{Guyer}},
  \bibinfo{author}{\bibfnamefont{P.~A.} \bibnamefont{Johnson}},
  \bibinfo{author}{\bibfnamefont{C.}~\bibnamefont{Marone}}, \bibnamefont{and}
  \bibinfo{author}{\bibfnamefont{J.}~\bibnamefont{Carmeliet}},
  \bibinfo{journal}{Phys. Rev. E.} \textbf{\bibinfo{volume}{89}},
  \bibinfo{pages}{042204} (\bibinfo{year}{2014}).

\bibitem[{\citenamefont{Capozza et~al.}(2009)\citenamefont{Capozza, Vanossi,
  Vezzani, and Zapperi}}]{capozzaprl2009}
\bibinfo{author}{\bibfnamefont{R.}~\bibnamefont{Capozza}},
  \bibinfo{author}{\bibfnamefont{A.}~\bibnamefont{Vanossi}},
  \bibinfo{author}{\bibfnamefont{A.}~\bibnamefont{Vezzani}}, \bibnamefont{and}
  \bibinfo{author}{\bibfnamefont{S.}~\bibnamefont{Zapperi}},
  \bibinfo{journal}{Phys. Rev. Lett.} \textbf{\bibinfo{volume}{103}},
  \bibinfo{pages}{085502} (\bibinfo{year}{2009}).

\bibitem[{\citenamefont{Giacco et~al.}(2012)\citenamefont{Giacco, Lippiello,
  and Ciamarra}}]{giaccopre}
\bibinfo{author}{\bibfnamefont{F.}~\bibnamefont{Giacco}},
  \bibinfo{author}{\bibfnamefont{E.}~\bibnamefont{Lippiello}},
  \bibnamefont{and} \bibinfo{author}{\bibfnamefont{M.~P.}
  \bibnamefont{Ciamarra}}, \bibinfo{journal}{Phys. Rev. E}
  \textbf{\bibinfo{volume}{86}} (\bibinfo{year}{2012}).

\bibitem[{\citenamefont{Ciamarra et~al.}(2010)\citenamefont{Ciamarra,
  Lippiello, Godano, and de~Arcangelis}}]{prl2010}
\bibinfo{author}{\bibfnamefont{M.~P.} \bibnamefont{Ciamarra}},
  \bibinfo{author}{\bibfnamefont{E.}~\bibnamefont{Lippiello}},
  \bibinfo{author}{\bibfnamefont{C.}~\bibnamefont{Godano}}, \bibnamefont{and}
  \bibinfo{author}{\bibfnamefont{L.}~\bibnamefont{de~Arcangelis}},
  \bibinfo{journal}{Phys. Rev. Lett.} \textbf{\bibinfo{volume}{104}},
  \bibinfo{pages}{238001} (\bibinfo{year}{2010}).

\bibitem[{\citenamefont{Ciamarra et~al.}(2011)\citenamefont{Ciamarra,
  Lippiello, de~Arcangelis, and Godano}}]{epl2011}
\bibinfo{author}{\bibfnamefont{M.~P.} \bibnamefont{Ciamarra}},
  \bibinfo{author}{\bibfnamefont{E.}~\bibnamefont{Lippiello}},
  \bibinfo{author}{\bibfnamefont{L.}~\bibnamefont{de~Arcangelis}},
  \bibnamefont{and} \bibinfo{author}{\bibfnamefont{C.}~\bibnamefont{Godano}},
  \bibinfo{journal}{Europhys. Lett.} \textbf{\bibinfo{volume}{95}},
  \bibinfo{pages}{54002} (\bibinfo{year}{2011}).

\bibitem[{\citenamefont{Ciamarra et~al.}(2009)\citenamefont{Ciamarra,
  de~Arcangelis, Lippiello, and Godano}}]{modphys2009}
\bibinfo{author}{\bibfnamefont{M.~P.} \bibnamefont{Ciamarra}},
  \bibinfo{author}{\bibfnamefont{L.}~\bibnamefont{de~Arcangelis}},
  \bibinfo{author}{\bibfnamefont{E.}~\bibnamefont{Lippiello}},
  \bibnamefont{and} \bibinfo{author}{\bibfnamefont{C.}~\bibnamefont{Godano}},
  \bibinfo{journal}{Int. J. Mod. Phys. B} \textbf{\bibinfo{volume}{23}},
  \bibinfo{pages}{5374} (\bibinfo{year}{2009}).

\bibitem[{\citenamefont{B.~Andreotti and Pouliquen}(2013)}]{andreotti}
\bibinfo{author}{\bibfnamefont{Y.~F.} \bibnamefont{B.~Andreotti}}
  \bibnamefont{and}
  \bibinfo{author}{\bibfnamefont{O.}~\bibnamefont{Pouliquen}},
  \emph{\bibinfo{title}{Granular Media: Between Fluid and Solid}}
  (\bibinfo{publisher}{Cambridge University}, \bibinfo{address}{Cambridge},
  \bibinfo{year}{2013}).

\end{thebibliography}
\end{document}